\documentstyle[epsfig,12pt]{article}
\begin{document}

\title{ {\bf About the parity-non-conserving asymmetry in $\vec{n} + p 
\rightarrow d+\gamma$ } }

\author{Bertrand DESPLANQUES$^a$ \thanks{ E-mail address:  
desplanq@isn.in2p3.fr}\\
$^a$Institut des Sciences Nucl\'eaires (UMR CNRS/IN2P3-UJF),  
\\ F-38026 Grenoble Cedex, France \\}

\maketitle

\begin{abstract}
The parity-non-conserving (pnc) asymmetry in $\vec{n} + p \rightarrow 
d+\gamma$ at thermal energies has recently been calculated using effective 
field-theory methods. A comparison of this calculation with much more elaborate 
calculations performed in the 70's is made. This allows one to assess the 
validity of this new approach as presently used. It is found to overshoot the 
almost exact calculations by a factor close to 2 for the contribution involving 
the $^3S_1$ component 
of both the initial and final states. This is much larger than anticipated by 
the authors. This discrepancy is analyzed and found to originate from the 
over-simplified description of the deuteron and capture states which underlies 
the new approach. The claim that earlier determinations of the sign would be in 
error is also examined. It is found that the sign discrepancy is most probably 
due, instead, to the fact that the pion-nucleon interaction referred to by the 
authors corresponds to a parity-non-conserving potential with a sign opposite 
to what is currently used. Some estimates and constraints relative to the pnc 
$\pi$NN coupling, $h^1_{\pi}$, which the above asymmetry is dependent on, are 
reviewed. Further details are given in an Appendix.
\end{abstract} 
\vspace{1cm}
\section{Introduction}
At the time where a new measurement of the parity-non-conserving (pnc) 
asymmetry in radiative  neutron-proton capture at low energy is in preparation 
at LANSCE \cite{LANS}, intending to improve upon a previous one at 
ILL \cite{ILL}, it is appropriate to wonder about the validity of theoretical 
estimates of the effect. This one is expected to be dominated by the 
contribution of the pnc pion-exchange force and its measurement should 
therefore provide unique information on the pnc $\pi$NN coupling, $h^1_{\pi}$, 
which it 
strongly depends on. This coupling constant is an important ingredient of pnc NN 
forces and its determination represents a major goal in the field (see 
\cite{DESP} for a review). This program supposes that theoretical uncertainties 
related to the description of the deuteron and np capture states are small 
enough so that the coupling can be reliably extracted from the measurement. 
By examining  studies performed in the 70's \cite{TADI,DESP1,MACK,GARI} and 
later \cite{MORI}, some values with different signs and magnitudes can be 
found. Yet, it was  concluded that the estimates of the asymmetry were rather 
independent of the strong interaction model \cite{MISS} but not much detail was 
published on the relevance of some ingredients (see below).  When correct ones 
are used, the relation between the pnc asymmetry to be measured, $A_{\gamma}$, 
and the above coupling, $h^1_{\pi}$, is essentially given by: 
\begin{displaymath}
 A_{\gamma} = -0.11 \, h^1_{\pi}.
\end{displaymath}
From the comparison of results obtained with reasonable strong interaction 
models, an uncertainty of about 5\% could be ascribed to the above value. This 
one is slightly larger than the first estimate by Danilov \cite{DANI},  
$A_{\gamma} = -0.08 \, h^1_{\pi}$, which was obtained on the basis of an 
approach directly relying on nucleon-nucleon phase shifts, but neglecting the 
effect of the tensor force.

Recently, a new calculation based on a different approach, namely an effective 
field theory \cite{SAVA}, came with a larger value,  $ A_{\gamma} = +0.17 \, 
h^1_{\pi}$, and furthermore an opposite sign. An uncertainty of about 30\% was 
ascribed to this value by the authors. This one could thus be compatible with 
the previous one, but it could also be twice as large. Apart from the fact that 
the discrepancy invalidates at first sight the argument that the coupling, 
$h^1_{\pi}$, could be determined unambiguously from the measurement 
of $ A_{\gamma}$, it is important to determine how reliable is the new 
calculation and how it should be corrected when improvements are made. Notice 
that values of the asymmetry with a size comparable to the above one may be 
found in the literature but they  correspond to a singlet scattering length 
inappropriate for the neutron-proton system.

The sign of the effect is another issue which deserves attention. It is the 
first information that one can hope to extract from an experiment, prior to the 
size itself. When calculations of the asymmetry $ A_{\gamma}$ were performed in 
the early 70's, it was believed that the sign of the pnc coupling, $h^1_{\pi}$, 
could not be determined theoretically, partly explaining the existence in the 
literature of relations of $ A_{\gamma}$ to $h^1_{\pi}$ with different signs 
\cite{MISS}. Since a theoretical determination of this sign has been made  
relatively to the parity conserving one, $g_{\pi NN}$, (this is the information 
that matters) \cite{DESP2}, it is essential to establish what is the correct 
sign for the relation of $ A_{\gamma}$ to $h^1_{\pi}$.

It this paper, we intend to compare the recent calculation to earlier ones, 
which, because they were quite sophisticated, can be used as a benchmark for 
testing  the applicability of the effective field-theory methods to the process 
under consideration.  At the same time, we discuss the points that are sources 
of discrepancy between calculations. After reminding a few ingredients required 
for the calculation of 
the pnc asymmetry of interest here, we show that the expression obtained in  
the new approach at the leading order identifies to a standard potential based 
calculation employing zero-range nucleon-nucleon forces (Sect. 2). In the third 
section, we discuss the size of the effect in relation with a better 
description of the nucleon-nucleon states. This is partly done by using 
analytical calculations, which has the advantage to show how the result changes 
while improvements are made. Section 4 is devoted to the sign of the 
effect. Details are given, allowing one to check its derivation. We, in 
particular, discuss the independence of this sign with respect to conventions 
that appear at intermediate steps of the calculation. In the fifth 
section, we present a short critical review of what is known, both theoretically 
and phenomenologically, on the elementary pnc  $\pi$NN  coupling constant, 
$h^1_{\pi}$. More details about this coupling are given in an Appendix.
\section{Main ingredients entering the calculation of $h^1_{\pi}$}
Due to parity-non-conservation in nucleon-nucleon forces, the emission of 
photons in the thermal neutron-proton capture can evidence an asymmetry with 
respect to the neutron polarization. This one, $A_{\gamma}$, is defined by 
the angular dependence of the capture cross section on the angle, $\theta$, 
between the direction of the photon emission and the neutron polarization, 
$W(\theta) \propto (1+ A_{\gamma} \, {\rm cos}\, \theta)$. A simplified 
expression, which accounts for the dominant contributions of interest here, is 
given by: 
\begin{equation}
A_{\gamma}= - 2 \; \frac{\tilde{E}1}{M1}.
\end{equation}
In this formula, the factor 2 is a current one appearing in expressions for an 
asymmetry. The sign is well defined but supposes that appropriate  conventions 
are used to calculate the matrix elements, $\tilde{E}1$ and $M1$. This will be 
detailed in the fourth section together with the spin-isospin algebra relative 
to the corresponding operators as well as to the pnc pion-exchange potential. 
The calculation of the two matrix elements are performed from the standard 
expressions for the electric and magnetic, $\Delta J=1$, transition operators. 
For the first one, we rely on the Siegert theorem, which allows one to sum up 
various contributions implied by gauge invariance, including those pertaining to 
the description of the pnc interaction. The sign difference which depends on 
whether the photon is absorbed or emitted is taken care of. For simplicity, 
factors that cancel out in the ratio given by Eq. (1) are ignored. 

The matrix element $M1$ of interest here involves the overlap of the wave 
functions of the deuteron state and the $^1S_0$ scattering state, which is the 
dominant one and provides most of the contribution to the capture cross 
section. It is given by:
\begin{equation}
M1= (\frac{\mu_p- \mu_n}{2 M_N}) \;\int dr\, r^2 \; \psi_d(r) \,\psi_s(r). 
\end{equation}
The matrix element, $\tilde{E}1$, involves a transition between the deuteron 
state and the $^3S_1$ scattering state. Its value, which is zero if parity is 
conserved, depends on the odd-parity component admixed to either state by the 
pnc nucleon-nucleon force. Capture from the $^3S_1$ state is not large, but it  
appears here because a state with a polarized neutron necessarily results 
from its coherent superposition with a $^1S_0$ state. The expression of the 
matrix element is given by:
\begin{equation}
\tilde{E}1=\frac{1}{3} \, \int dr \, r^2 \; \left(\psi_d(r) \, r \, 
\tilde{\psi}_t(r) 
 - \tilde{\psi}_d(r) \, r \, \psi_t(r)\right),
\end{equation}
where the factor $\frac{1}{3}$ comes from integrating over angles the angular 
dependence due to both the dipole operator and the parity admixed component 
with the result 
$$\int \frac{d\Omega}{4\pi} \, r^i r^j= \, \frac{1}{3} \, r^2 \, \delta^{ij}.$$
The minus sign in Eq. (3) comes from taking the complex conjugate of the 
complete deuteron wave function whose parity admixture contains the extra 
imaginary number, i (which combines with a i in the electric dipole operator to 
produce a real number). For simplicity, the admixture of a $^3D_1$ component to 
the $^3S_1$ one has been omitted in the above equation (as it was in 
\cite{SAVA}) but it is accounted for in the most elaborate calculations whose 
results are reminded below. For the calculation of interest here, 
its effect can be accounted for by replacing the current $^3S_1$-wave component, 
$u(r)$, by the combination involving the D-wave component, 
$u(r)+\frac{1}{\sqrt{2}} \, w(r)$. As to the quantities, $\tilde{\psi}_d(r)$ 
and $\tilde{\psi}_t(r)$, they respectively describe the parity admixed 
components to the deuteron and the  $^3S_1$ scattering states. As shown by 
Danilov \cite{DANI}, these components involve a pnc transition from a $^3S_1$  
to a $^3P_1$ state (the electric dipole operator does not change the spin for 
its dominant contribution) and, thus, implies a $T=0$ to a $T=1$ transition. 
This can only arise from an isovector pnc force.

The isovector pnc force is expected to be largely dominated by the pion-exchange 
contribution, due to the relatively low mass of the pion and, 
consequently, its somewhat long range. The expression for this force, first 
given in \cite{MACK0}, is currently written down as follows \cite{DDH, 
ADEL,DESP}:
\begin{equation}
V_{\pi}= i \, \frac{g_{\pi NN} \, h^1_{\pi}}{ 2\sqrt{2} } \; 
(\vec{\tau}_1 \times \vec{\tau}_2)^z \, (\vec{\sigma}_1 + \vec{\sigma}_2)
\left[\frac{\vec{p}_1-\vec{p}_2}{2M_{N}}, \frac{e^{-m_{\pi} r}}{4\pi \,r} 
\right],
\end{equation}
where $g_{\pi NN}$ represents the strong $\pi$NN coupling constant, which is 
taken positive here as usually done. Contrary to the meson-nucleon 
interaction, the above 
expression of the pnc potential is essentially free of any convention. The sign 
of $V_{\pi}$ is therefore determined by the sign of $ h^1_{\pi}$ (see Sect. 4 
for more details).

Knowing the expression of the pnc force, one can now  determine the parity 
admixed components it produces in the total deuteron or $^3S_1$ scattering 
states, which we write as:
\begin{equation}
|\Psi>_{d,t}= \left( \psi_{d,t}(r)+... +i \, \frac{\tau^z_p - 
\tau^z_n}{2} \, (\vec{\sigma}_p+\vec{\sigma}_n).
\frac{\vec{r}_p-\vec{r}_n}{|\vec{r}_p-\vec{r}_n|} \,   
\tilde{\psi}_{d,t}(r) \right)|S=1>,
\end{equation}
where the dots stand for the tensor component we omit for simplicity. From now 
on, particles 1 and 2 are ascribed to the proton and neutron  respectively, but
we nevertheless keep track of the isospin factor, equal to 1, to evidence the 
symmetry between the two particles. As the pnc force is a small perturbation, 
the expressions of the parity admixed components, 
$\tilde{\psi}_d(r)$ and $\tilde{\psi}_t(r)$, can be obtained using the full 
Green's function projected on $l=1$ angular momentum states. For the deuteron 
state, one gets:
\begin{equation}
\tilde{\psi}_d(r)= \int dr' \, r'^2 \; G^{l=1}_E(r,r')
  \frac{g_{\pi NN} \, h^1_{\pi}}{ 4\pi M_N \sqrt{2} } \;
  \frac{e^{-m_{\pi} r'}}{r'^2}(1+m_{\pi} r') \; (\psi_d(r')+...),
\end{equation}
where $E=- |E_d|$. A similar expression holds for the scattering state with the 
appropriate replacements. Spin and isopin dependences have been factored out so 
that Exp. (1) (together with Exps. (2, 3, 6)) offers no ambiguity as to its 
sign once the sign of $ h^1_{\pi}$ is given (intrinsic phases relative to the 
deuteron or scattering states cancel out).

A first but unrealistic estimate of the asymmetry, $A_{\gamma}$, can be obtained 
by using the following wave functions corresponding to a zero-range force 
together with the free Green's function projected on the $l=1$ space:
\begin{equation}
\psi^0_s(r)= 1- \frac{a_s}{r}, \;\;\;\;
\psi^0_t(r)= 1- \frac{1}{\alpha r}, \;\;\;\;
\psi^0_d(r)= \sqrt{2 \alpha} \; \frac{e^{-\alpha r}}{r},
\end{equation}
\begin{eqnarray}
G^{l=1}_{E=0} (r,r') =  -M_N  \, \frac{1}{\,r^2} \, \left(\frac{r'}{3} \;\; 
\theta(r - r') \right) +..., \hspace{5cm} \nonumber \\
G^{l=1}_{E=-|E_d|}(r,r')  = 
-M_N \, \frac{e^{-\alpha r} \,(1+ \alpha r)}{r^2} \hspace{6.5cm} \nonumber \\
  \left( \frac{e^{-\alpha r'} \, (1+ 
\alpha r')-e^{\alpha r'} \,(1- \alpha r')}{2  \alpha^3 \, r'^2 }    
\;\; \theta(r - r') \right)+....
\end{eqnarray}
In these equations, $\alpha$ is related to the deuteron binding energy, $\alpha 
= \sqrt{M_N |E_d|}$. The expression of the Green's function, only given for  
$r'<r$, allows one to precise factors.
 
The calculation of integrals in Eqs. (2, 3, 6) can be performed analytically 
with the result:
\begin{equation}
A^0_{\gamma}= -\frac{ g_{\pi NN} \, h^1_{\pi}\, M_N \sqrt{2}}{3 \pi \,(\mu_p- 
\mu_n)(1- \alpha a_s)}
\left[ \frac{m_{\pi}}{(m_{\pi}+\alpha)^2}
 + \frac{m_{\pi}^2}{\alpha^2 (m_{\pi}+\alpha)} - \frac{m_{\pi}^2}{2\alpha^3} 
\, {\rm log}(1+\frac{2\alpha}{m_{\pi}}) \right].
\end{equation}
This expression is identical to that one obtained by Kaplan et al. when using 
the effective field theory at the leading order. In view of the extreme 
simplicity of the wave functions it corresponds to, the result can  be easily  
improved. By using more realistic wave functions, one can provide insight on 
the corrections that should be applied to the estimate of $A_{\gamma}$ made by 
these authors. Let's notice that, due to different inputs concerning mainly the 
strong $\pi NN$ coupling, the magnitude of the asymmetry given by the above 
equation is slightly larger than what they obtain. The 
discrepancy has not much physical relevance but is important in order to assess 
the precise validity of the approach. Thus, the value to be compared with more 
elaborate estimates presented below is $-0.187\,h^1_{\pi}$ (corresponding to 
$\frac{g^2_{\pi NN}}{4\pi}=14.4,\, m_{\pi}=0.7 \,{\rm fm}^{-1},\, M_N=4.76 
\,{\rm fm}^{-1}, \,\mu_p- \mu_n=4.706,\, \alpha=0.232\,{\rm fm}^{-1},\, 
a_s=-23.7\,$fm).

\section{Size of the effect}
The size of the pnc asymmetry, $A_{\gamma}$, has been calculated in many papers 
\cite{DANI,TADI,DESP1,MACK,GARI,MORI}. They evidence discrepancies depending on 
whether results are corrected for the fact that most realistic models 
used in the past were not providing the right neutron-proton scattering length 
for the $^1S_0$ state. These models were assuming isospin symmetry, often 
predicting a scattering length of about $-17\,$fm. This represents the main 
source of difference, roughly a factor 1.4 ($\simeq \frac{-23.7}{-17}$). 
Actually, this can be remedied by taking the value of the regular 
transition amplitude, $M1$, from the measured capture cross section, which 
offers the further advantage to also account for a 5\% correction due to 
meson-exchange currents. Results also depend on whether interaction in P 
waves or effects due to the tensor force are considered. When the main
ingredients mentioned above are accounted for \cite{MISS}, the uncertainty is 
due to the strong interaction models themselves and the precise  way to correct 
for their deficiencies. Asymmetries calculated on the same footing with the 
Hamada-Johnston, Reid-soft-core and de Tourreil-Sprung potential models for 
instance were found to be equal to $-0.109\,h^1_{\pi}$, $-0.114\,h^1_{\pi}$ and 
$-0.107\,h^1_{\pi}$  respectively \cite{DESP1}. In all cases, the effect of the 
repulsion in the $^3P_1$ state ($\simeq 25-30\%$) is partly compensated by an 
effect of about 20\% due to the admixture of a $^3D_1$ component to the 
$^3S_1$ one. Careful examination also indicates that part of the 
differences has a well determined origin: a slightly too large binding energy 
for the deuteron in the case of the Hamada-Johnston model and a 2\% difference 
for the triplet scattering length for the two other models. The discrepancies 
can therefore be made smaller by using improved models of the NN interaction.

The insensitivity to the strong interaction model was confirmed later on by 
using the Paris potential \cite{MORI}, which is a more up to date model, 
including an accurate treatment of the two pion-exchange contribution. Prior to 
the comparison, the result has to be corrected for a wrong $^1S_0$ scattering 
length however.

The above values have a magnitude smaller than what was obtained by Kaplan et 
al., $-0.187\,h^1_{\pi}$ with our inputs (see remark end of previous section). 
The discrepancy 
can be easily ascribed to the oversimplified wave functions, which the Kaplan 
et al.'s calculation corresponds to (Eq. 9). With respect to these 
ones,  Hulth\'en wave functions (corresponding to a separable Yamaguchi 
potential) represent a significant improvement although they 
are not good ones in view of the present standard. They offer the 
advantage of providing an analytic expression for the pnc asymmetry, 
$\tilde{E}1$, which can be used to estimate the effect of the finite range of 
the strong interaction neglected in the above calculation. It is given by:

\begin{eqnarray}
A^1_{\gamma}= -\frac{ g_{\pi NN} \, h^1_{\pi}\, M_N \sqrt{2}}{3 \pi \,(\mu_p- 
\mu_n) \; \left(1- (\frac{\alpha}{\eta_t})^2 -\alpha a_s(1- 
\frac{\alpha}{\eta_t}
-\frac{\alpha}{\eta_s+\alpha} + \frac{\alpha}{\eta_s+\eta_t}) \right) } 
\hspace*{4cm} 
\nonumber \\
 \times \left[ \frac{m_{\pi}}{(m_{\pi}+\alpha)^2}
   + \alpha a_t \, \left( \frac{m_{\pi}^2}{\alpha^2 (m_{\pi}+\alpha)} - 
  \frac{m_{\pi}^2}{2\alpha^3} 
  \, {\rm log} (1+\frac{2\alpha}{m_{\pi}} ) \right) 
\right. \hspace*{4cm} \nonumber \\ \left. 
   - \frac{m_{\pi}}{(m_{\pi}+\eta_t)^2}                                            
    -\frac{\eta_t^2-\alpha^2}{2\eta_t(m_{\pi}+\eta_t)^2}
\right. \hspace*{8cm} \nonumber \\ \left.
  - \alpha a_t 
 \left(  \frac{2m_{\pi}^2-2\eta_t^2-\alpha m_{\pi}+\alpha \eta_t}{4 \alpha^2    
                (m_{\pi}+\eta_t)}
   + \frac{2m_{\pi}^2-2\eta_t^2+\alpha m_{\pi}-\alpha \eta_t+\alpha^2}{4 
  \alpha^2     (m_{\pi}+\eta_t+\alpha)}
\right. \right. \hspace*{3cm} \nonumber \\ \left. \left.
  -\frac{2 m_{\pi}^2-2 \eta_t^2 + \alpha^2}{2 \alpha^3} 
    {\rm log}  (\frac{m_{\pi}+\eta_t+\alpha}{m_{\pi}+\eta_t})  \right) 
\right. \hspace*{5cm} \nonumber \\  \left.
  + \frac{\alpha^2 a_t}{(\eta_t+\alpha)^2} 
  \left(\frac{2m_{\pi}^2+\eta_t^2 +m_{\pi} \eta_t-\alpha m_{\pi}  
  -\alpha^2}{4(\eta_t-\alpha)(m_{\pi}+\eta_t+\alpha)}
  - \frac{m_{\pi}^2}{2(\eta_t-\alpha)^2} {\rm log}  
  (\frac{m_{\pi}+\eta_t+\alpha}{m_{\pi}+2\alpha}) \right) 
\right. \hspace*{1cm} \nonumber \\ \left. 
  - \frac{\alpha^2 a_t}{\eta_t^2} 
  \left( \frac{2m_{\pi}^2+\eta_t^2+m_{\pi}\eta_t}{4\eta_t(m_{\pi}+\eta_t)}
   -\frac{m_{\pi}^2}{2\eta_t^2}  {\rm log}  
  (\frac{m_{\pi}+\eta_t}{m_{\pi}})\right) 
\right.\hspace*{3cm}   \nonumber \\ \left. 
  - \frac{\alpha^2 a_t}{(\eta_t+\alpha)^2} 
  \left(\frac{2m_{\pi}^2-2\eta_t^2+ m_{\pi}\eta_t -\alpha m_{\pi} +2\alpha       
   \eta_t }{4(\eta_t-\alpha)(m_{\pi}+2\eta_t)}
  - \frac{m_{\pi}^2-\eta_t^2+\alpha^2}{2(\eta_t-\alpha)^2} {\rm log}  
  (\frac{m_{\pi}+2\eta_t}{m_{\pi}+\eta_t+\alpha}) \right) 
\right. \nonumber \\ \left.   
  + \frac{\alpha^2 a_t}{\eta_t^2} 
  \left( \frac{2m_{\pi}^2-2\eta_t^2+m_{\pi}\eta_t}{4\eta_t(m_{\pi}+2\eta_t)}
   -\frac{m_{\pi}^2-\eta_t^2}{2\eta_t^2}  {\rm log}  
  (\frac{m_{\pi}+2\eta_t}{m_{\pi}+\eta_t}) \right)   \right]. \hspace*{3cm}    
\end{eqnarray}
In the limit of a zero-range force ($\eta_t \rightarrow \infty,\, 
\alpha a_t \rightarrow 1$), the front factor together with the first line in the 
above equation allow one to recover Eq. (9), while the other terms vanish as 
$\eta_t^{-1}$ or $\eta_t^{-2}$. For the value that $\eta$ is expected to take in 
the case of a finite range force ($\eta_t= 1.386\,{\rm fm}^{-1}$ corresponding 
to $a_t=5.42\,{\rm fm}$ and $\alpha=0.232\,{\rm fm}^{-1}$, and 
$\eta_s=1.203\,{\rm fm}^{-1}$), one gets a reduction by a factor 1.33. 

Another analytical estimate of the effect can be obtained using the Danilov's 
approach \cite{DANI}. This one, which relies on dispersion relations, only uses 
the knowledge of NN scattering phase shifts. It has been shown to give a good 
account of pnc effects in the neutron-proton radiative capture process, the 
uncertainty being of the order of 20-25\% in absence of accidental cancellation 
\cite{EPST}. In this approach, the whole effect is factorized into two parts 
involving the electric dipole operator and the pnc NN force, which respectively 
are sensitive to the long and medium plus short range description of NN states. 
The pnc effect is thus determined by the strength of the pnc neutron-proton 
scattering amplitude, $M_NC$, whose expression is given by:
\begin{equation}
M_NC= - \frac{g_{\pi NN} \, h^1_{\pi}}{4 \pi \sqrt{2}} \frac{ M_N}{a_t}
 F_t(0) \; \int_{-\infty}^{ -\frac{m^2_{\pi}}{4} }dp'^2 \; 
(-\frac{m^2_{\pi}}{8 \, p'^6}) \; F^{-1}_t(p'^2).
\end{equation}
where
\begin{equation}
F^{-1}_t(p^2)=(1+\frac{p^2}{\alpha^2}) \;\; {\rm exp} \left( \frac{p^2}{\pi}
\int_0^\infty \frac{dp'^2 (\delta_t (p'^2)-\pi) }{p'^2 \, (p^2-p'^2+i\epsilon)} 
\right),
\;\;(\delta(0)=\pi).
\end{equation}  
Inserting in this equation the standard low energy effective range 
parametrization for phase shifts, 
$\frac{p}{{\rm tg} \, \delta_t}= -\frac{1}{a_t} +\frac{r_t\, p^2}{2}$, one 
gets:
\begin{equation}
F^{-1}_t(p^2)=\frac{ 1-\frac{\sqrt{-p^2}}{\alpha} 
}{ 1+\frac{\alpha a_t r_t \sqrt{-p^2}}{2} } \;\;({\rm for} \, p^2  < 0),
\end{equation}  
and from it:
\begin{eqnarray}
M_NC & = & - \frac{ g_{\pi NN} \, h^1_{\pi} }{ 4 \pi \sqrt{2} } 
\frac{ M_N}{a_t} \int_{ \frac{m_{\pi}}{2} }^{\infty} dq \, \frac{m^2_{\pi}}{4 
\, q^5} \, \frac{ 1-\frac{q}{\alpha} }{ 1+ \frac{\alpha  a_t r_t \, q }{2} }  \\
 & = & \frac{ g_{\pi NN} \, h^1_{\pi} }{ 4 \pi \sqrt{2} } 
 \frac{ M_N}{\alpha a_t m_{\pi}}
\left(  (1+\frac{2\alpha}{m_{\pi}} \, x)
(\frac{2}{3}- x + 2x^2 -2x^3 {\rm log}(1+\frac{1}{x})) -\frac{\alpha}{m_{\pi}}  
\right), \nonumber
\end{eqnarray}
with $x=\frac{\alpha \, a_t \, r_t \, m_{\pi}}{4}$.
A numerical estimate of the effect of the range is immediately obtained by 
comparing values of $M_NC$ calculated with $r_t=\, 1.765\,$fm and $r_t=\,0$. A 
reduction by a factor 1.6 is thus found in the case of the scattering state. 
This is partly reduced by the fact that the scattering length $a_t$ is 
simultaneously enhanced, the real reduction factor being closer to 1.3. 

The apparent independence of expressions (11-14) of strong interaction models 
(it is sufficient to know the phase shifts) is partly misleading. It is likely 
that the off-shell effects that the underlying approach neglects have to be 
accounted for by extra contributions to the dispersion relation, Eq. (11). 

Estimates due to a better description of the S states given above do not take 
into account the repulsion at short distances. From the comparison with results 
from realistic NN interaction models, an extra reduction by a factor $\simeq 
1.15-1.20 $ is found for this part. Some reduction by a factor $\simeq 
1.25-1.30$ is also due to the repulsion in the $^3P_1$-state. Thus the total 
reduction of $A_{\gamma}$ with respect to the value  $-0.187\,h^1_{\pi}$ amounts 
to a factor $\simeq 1.9-2.0$. The size of the effect  
is larger than the 30\% referred to in ref. \cite{SAVA}. With this respect, we 
would like to mention that this uncertainty is almost reached by the effect of 
neglecting the finite range of the interaction. The value of the product 
$\alpha \, a_t$, which is set to 1 in the zero-range limit, is actually larger 
by a factor 1.26. The deuteron asymptotic normalization, $\sqrt{2 \alpha}$, is 
found smaller than the experimental one, 0.884 fm$^{-\frac{1}{2}}$ by a factor 
1.30. As these discrepancies concern long range properties, one should not be 
surprised by the significantly larger discrepancy we found for the pnc effect. 
This one depends on the short range properties that  are much less well 
described by the present use of effective field-theory methods. 

The agreement obtained for the total capture cross section was considered by 
Kaplan et al. as a support for their approach \cite{SAVA}. In view of the 
previous comments, it looks somewhat accidental. It results from a cancellation 
of two effects: a long range contribution that is too small and a shorter range 
contribution (further enhanced for the pnc case) that is too large. 

\section{Sign of the effect}
Signs in the field of parity violation are an important information but have 
not always received the attention they deserve. The size of the measured pnc 
asymmetry in pp scattering at low energy has been reproduced in many papers for 
instance. A careful examination nevertheless showed that ten of them  were  
actually missing the sign \cite{DESP3}. In view of this, we here precise 
ingredients that enter into the determination of the sign of the pnc asymmetry, 
$A_{\gamma}$, together with its expression. They concern the pnc potential, 
$V_{\pi}$, the electromagnetic transition operators, and the calculation of the 
asymmetry itself.

The determination of the sign of the pion-exchange pnc force, Eq. (4), has been 
made for the first time in ref. \cite{DESP2}. It makes sense as far as the 
expression of Pauli matrices and the relation $\vec{p}=-i\vec{\nabla}$  are 
universally accepted. The sign is then related to the sign of the product 
of the strong and weak $\pi$NN couplings. Individually, the sign of each of 
them depends on many conventions: intrinsic phase of the meson field, 
definition 
from an Hamiltonian or from a Lagrangian, expression of the $\gamma$ matrices, 
especially the $\gamma_5$ matrix which makes the difference between a pc and a 
pnc interaction, spin-$\frac{1}{2}$ fields referring to particles or 
anti-particles, metric, ... While it is possible to get rid of all these 
conventions with some care, this is not sufficient to  determine the sign of the 
product of the couplings. The last step is achieved by noticing that the 
calculation of the pnc coupling, $h^1_{\pi}$, using PCAC and current algebra, 
is involving a factor proportional to the strong coupling, hence a dependence 
of $g_{\pi NN} \, h^1_{\pi}$ on  the quantity $g^2_{\pi NN}$, which is always 
positive \cite{DESP2}. The sign of the three types of contributions considered 
in DDH (sum rule, ``parity admixture'' into the wave function and factorization; 
see Fig. 1c, 1b, 1a in the Appendix) 
could thus be determined. Taking as positive the strong coupling appearing in 
the pnc potential, Eq. (4), as done in DDH and subsequent papers 
\cite{DDH,DESP},  the estimated weak coupling $h^1_{\pi}$ and its range given in 
these works correspond to a positive value.

The electric $E1$ and magnetic $M1$ operators are a possible source of sign 
mistake. They can be derived from the electromagnetic Hamiltonian, which may be 
written as:
\begin{equation}
H_{el}= e \sum_j \left( ( \frac{ (\vec{ p}_i + \vec{ p}_f) \, .\, \vec{ 
\epsilon}}{2\,M_N})_j \, 
(\frac{1+\tau^z}{2})_j
\pm i (\frac{\mu_p}{2M_N}\, 
\frac{1+\tau^z}{2}+\frac{\mu_n}{2M_N}\,\frac{1-\tau^z}{2})_j
\; \vec{ \sigma}_j \, . \, \vec{ q} \times \vec{ \epsilon} \, \right) +...
\end{equation}
where the dots account for meson-exchange currents  required to ensure 
gauge invariance. The photon polarization is represented by the vector 
$\vec{ \epsilon}$ and its momentum (energy) by $\vec{ q} \, (q^0)$. An 
alternative expression, which is of more direct usefulness for the present work, 
relies on the Siegert theorem to account for the dominant 
contributions due to exchange currents (represented by dots below), 
$$\sum_j (\frac{ \vec{ p}  \, .\, \vec{ \epsilon}}{2\,M_N})_j \, 
(\frac{1+\tau^z}{2})_j+.... 
=i \, \frac{1}{2} \, \left[ H_{pc}+H_{pnc}, \sum_j \vec{r}_j \, 
(\frac{1+\tau^z}{2})_j  \right] .$$ 
It is given by:
\begin{equation}
H_{el}= e \left(\pm i \, \frac{1}{2} \, q^0 \, (\vec{ r_p}-\vec{ r_n}).\vec{ 
\epsilon} \pm i \, \frac{1}{2} \,  \frac{\mu_p-\mu_n}{2M_N}   
( \vec{ \sigma_p}-\vec{ \sigma_n} ) \, . \, \vec{ q} \times \vec{ \epsilon}   
\right) \, \frac{\tau^z_p -\tau^z_n}{2}+....
\end{equation}
where the dots, here and below, represent contributions that are irrelevant for 
our purpose. Different signs appear in Eq. (16) as well as Eq. (15), depending 
on whether an absorption or an emission of the photon is considered (or on 
whether one calculates the matrix element $<i|H|f>$ or $<f|H|i>$). 
The present ones are given for an emission (radiative capture) and respectively 
correspond to $<i|H|f> \; (<np|H|d\gamma>)$ and $<f|H|i> \; (<d\gamma|H|np>)$ 
for the upper and lower sign. Either way leads to the same result, but we 
nevertheless emphasize the point as the electric  dipole operator, $\vec{ r}$, 
in Eq. (16) has not always been used with the appropriate sign for the energy 
factor $q^0$, giving rise to different signs for the parity violating 
asymmetries \cite{DESP1,MISS}. 

Using the expression of the electromagnetic interaction, Eq. (16), the 
expression of the total wave function, Eq. (5), together with Eq. (6) for the 
parity admixed component, one obtains the transition matrix operator in terms 
of the matrix elements, $M1$ and $\tilde{E}1$, previously defined, Eqs. (2, 3):
\begin{equation}
T(n \, p \rightarrow d\, \gamma) \propto \left(
\pm i \, M1 \,   \,(\vec{ \sigma_p}-\vec{ \sigma_n})\, . \, \vec{ q} \times 
\vec{ \epsilon} +\tilde{E}1 \, (\vec{ \sigma_p}+\vec{ \sigma_n}) \, .
\,(q^0 \, \vec{ \epsilon} - \vec{ q} \, \epsilon^0) +....\right).
\end{equation}
The last term has been completed with a contribution from the 
charge density so that to emphasize its gauge invariance. Using the 
relation, $ i  \,(\vec{ \sigma_p}-\vec{ \sigma_n})= (\vec{ \sigma_p} 
\times \vec{ \sigma_n})\; P_{\sigma}$, one obtains an alternative expression.  
Less dependent on conventions as far as the relative sign of $M1$ and 
$\tilde{E}1$ contributions is concerned, it is given by:
\begin{equation}
T(n \, p \rightarrow d\, \gamma) \propto \left(
 M1 \, (\vec{ \sigma_p} \times \vec{ \sigma_n})\, . \, \vec{ q} \times 
 \vec{ \epsilon} +\tilde{E}1 \, (\vec{ \sigma_p}+\vec{ \sigma_n}) \, .
\,(q^0 \, \vec{ \epsilon} - \vec{ q} \, \epsilon^0) +....\right). 
\end{equation}

The angular distribution of the emitted photons with respect to the neutron 
polarization $\vec{n}$ is obtained by taking the matrix element of the 
transition operator, $T$, corresponding to a given neutron polarization along 
this direction, multiplying it by its complex conjugate and summing over all 
polarizations in the final states. This amounts to calculate:
\begin{eqnarray}
\sigma(\theta) \propto \sum_{\epsilon} {\rm Tr.} \;
( \;( \frac{ 1+\vec{\sigma_n }.\vec{n} }{2}) \;\; T \;\; 
(\frac{ 3+\vec{\sigma_p }.\vec{\sigma_n }}{4}) \;\; T^{+} \; ) \nonumber \\
\propto (M1^2 \, -2 \; M1 \; \tilde{E}1 \;\; \vec{n }. \hat{q} +...),
\end{eqnarray}
from which Eq. (1) immediately follows.
\section{Present estimates and constraints for the pnc coupling, $h^1_{\pi}$}
There are in the literature numerous papers dealing with the estimate of the 
pnc coupling, $h^1_{\pi}$, in the standard electro-weak interaction model 
\cite{REID,DESP2,WEIN,DDH,KHAT,DUBO,KAIS,HENL,HWAN,SAVA2,MEIS}. The most 
comprehensive work is still probably the DDH one \cite{DDH}. Contributions 
considered since then most often correspond to one of those estimated in this 
earlier work. While the improvement is not always obvious, at least, they 
provide a better estimate of the uncertainty that should be ascribed to each of 
them. Due to expected cancellations between different contributions, picking up 
one of them is however misleading. Some details and a discussion are given in 
ref. \cite{DESP} as well as in the Appendix. One of the features evidenced by 
the results, summarized in 
DDH, was their strong dependence on the factor, K, that accounts for strong 
interaction effects. Since then, due to improvements in determining the strong 
QCD coupling, the value of K has decreased from $K=6$ to $K=3$ for an energy 
scale of $\mu=1 \, ${\rm GeV}. It was also realized that there was some 
cancellation 
between the effect due to a dependence of the K factor on this (arbitrary) 
energy scale and that one for the quark masses \cite{LEUT}, with the result of 
making estimates of pnc meson-nucleon couplings less dependent on it, as it 
should be. The introduction of a contribution implying strange 
quarks in the nucleon could have the same effect for another contribution 
involving strange quarks in the hadronic weak interaction. Thus, an up-dated 
range for the coupling  would now be: $ 0 < h^1_{\pi} < 2.5 \, 10^{-7} $ (see 
\cite{ZHU} for an analysis of some uncertainties). Due to 
large cancellations in the total contribution involving u and d quarks (see 
Table 4 of ref. \cite{DESP} and Appendix), most of the contribution arises from 
the part of the 
weak interaction involving strange quarks. It is interesting to notice that a 
recent estimate of the same contribution \cite{MEIS} falls in the above range. 
Other estimates fall outside. While that one for the strange 
quark in ref. \cite{SAVA2} is not a serious source of concern in view of its 
very rough character, those based on QCD sum rules \cite{KHAT,HENL,HWAN}, which 
in any case ignore strange quarks, are. It is not clear which contribution they 
correspond to in the DDH picture. Let's finally mention that all studies give 
positive values for $h^1_{\pi}$. From the examination of the papers however, we 
did not get the certitude that the sign was carefully determined (assuming it 
was looked at). 

Phenomenological constraints on the coupling, $h^1_{\pi}$, have been recently 
discussed in the literature \cite{DESP, WILB}. To explain the recent 
measurement of the anapole moment in $^{133}Cs$ \cite{WOOD}, it has been 
suggested that a large value of the coupling would be required 
\cite{HAXT2,FLAM1}, exceeding the upper limit determined from the experiment 
performed at ILL in the past \cite{ILL} (  
$h^1_{\pi}= (1.4 \pm 4.4) \, 10^{-7}$, in absence of other contribution). An 
isoscalar pnc force enhanced in the nuclear medium could 
do as well with a coupling that would be smaller and essentially in agreement 
with what the absence of observed effect in $^{18}F$ indicates, $h^1_{\pi} \leq 
1.3\, 10^{-7}$ \cite{DESP}. It is remembered 
that the calculation of the nuclear pnc matrix element in this case can be 
checked by using information from first forbidden $\beta$ decay from $^{18}Ne$.  
Although it has not the same probing strength as in $^{18}F$, the absence of 
effect in $^{21}Ne$ and in $^{93}Tc$ also points to a small coupling if one 
discards it results from accidental cancellations \cite{DESP,DUMI}. Let's remind 
that the present theoretical contribution of the pion-exchange to the circular 
polarization of photons emitted in the transition, $\frac{1^{-}}{2} \rightarrow 
\frac{3^{+}}{2}$ in $^{21}$Ne is expected to be about $|P^{\pi}_{\gamma}|= 10^5 
\, |h^1_{\pi}| $, whereas the measurement puts the limit $P_{\gamma}= (0.8 
\pm 1.4) \times 10^{-3}$.
\section{Conclusion}
We have shown that, far from representing a progress, the recent calculation of 
the pnc asymmetry in $ \vec{n} + p \rightarrow d + \gamma $, based on using 
effective field-theory methods at leading order, is a step back to pre-historic 
times of the field. While this approach looks appealing at first sight, it 
evidences large discrepancies with  previous calculations. Realizing that these 
latter rely on much more elaborate descriptions of the deuteron and scattering 
states, they can be used to assess the validity of the new approach as used 
until now. 

From the comparison we made, the uncertainty of 30\% ascribed to it by the 
authors is reasonable for the calculation of observables that have a long 
range, such as magnetic or electric transitions. In the present case however, 
part of the effect involves a weak pnc interaction which has a shorter 
range. Examination of their expression for the asymmetry $A_\gamma$, which 
corresponds to an attractive zero-range force, shows that it behaves like 
$\frac{1}{m_{\pi}}$, while for reasonable potentials one rather expects 
$\frac{1}{m_{\pi}^2}$ (all other quantities being supposed to be small). This 
part of the calculation is overestimated by a factor of about 2. The 
discrepancy has its origin in the treatment of nucleon-nucleon correlations and 
especially their well known short range repulsive part, which prevents two 
nucleons to have an infinite probability to be close to each other (phase space 
factor put apart) as the present calculation by Kaplan et al. supposes. It 
decomposes into a factor roughly given by $(1+ c\, \frac{r_t \,m_{\pi}}{3}) 
\simeq 1.5-1.6$, representing the combined effect of a finite range attraction 
together with a short range repulsion for the force in the $^3S_1$ state ($r_t$ 
represents the effective range and $c \simeq 1.3$) and a factor $\simeq 
1.25-1.30$ representing the effect of long and short range repulsion in the 
$^3P_1$ states. The overestimate is actually smaller by a factor $\simeq 1.15$ 
due to a contribution  from the tensor force that is constructive in the present 
case, itself reduced by the correction of mesonic exchange currents to the 
regular transition, M1. While the new result by 
Kaplan et al. lets hope that the pnc asymmetry, $A_{\gamma}$, could be quite 
large in view of the error ascribed to the theoretical method, it thus appears 
that: 1) the error with respect to the actual value is underestimated, 2) this 
value is slightly outside the range they provided, in the lower part. 

The underestimate of the error indicates that the parameter which, in the 
approach, governs the expansion at different orders does not provide a good 
magnitude of their relative contributions. This conclusion is supported by the 
omission of the dominant contribution to the deuteron anapole moment in 
\cite{SPRI1} (corrected in \cite{SPRI2}) which suggests that the determination 
of what is relevant at some order is not so straightforward. This does not 
preclude the use of effective approaches for describing pnc effects however. 
Such approaches were initiated in the field by Danilov with a different 
foundation \cite{DANI} and extented to nuclei in \cite{MISS}. They can give a 
good account of pnc effects in a large variety of low energy processes.

The determination of the sign of the asymmetry requires some care, especially 
to get rid of numerous conventions that appear at different steps of the 
calculation. Assuming that the authors of ref. \cite{SAVA} have relied on 
commonly used 
definitions for writting down the expression of the pion-nucleon coupling, we 
found that the pion-exchange potential which it gives rise to has a sign 
opposite to the current one, thus producing an asymmetry with the sign 
different from that one generally referred to. After correcting their result 
for this different choice, the determination of the sign by these authors 
confirms the earlier determination, thus providing the independent check that 
they were calling for.

Concluding, there is a long way ahead before the effective field-theory 
approach to the calculation of the asymmetry $A_{\gamma}$ be able to compete 
with more elaborate calculations. For the pion-exchange contribution, these 
ones converge to the value $A_{\gamma}=-(0.11\pm 0.01)\, h^1_{\pi}$. The sign 
refers to the current expression of the $\pi$-exchange pnc potential and the 
error, estimated largely, results from the comparison of the most realistic 
descriptions of the process.
\vspace{2cm}

\appendix
 {\Large { \bf Appendix: More about $h^1_{\pi}$} } \\

\begin{figure}
\begin{center}
\mbox{\psfig{file=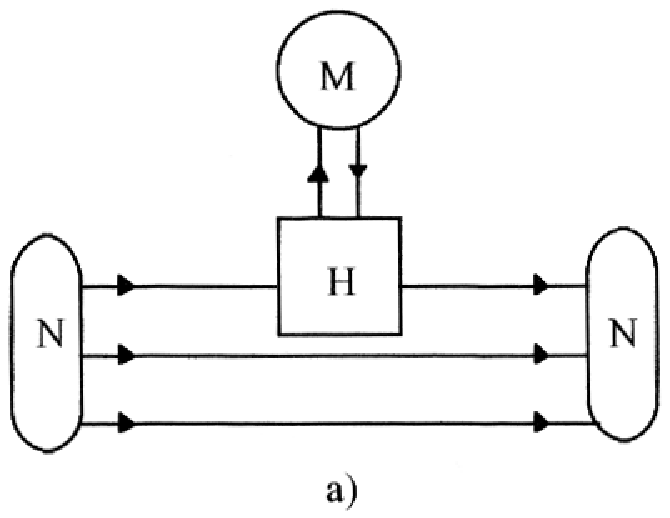,width=55mm}}
\vskip 0.5cm
\mbox{\psfig{file=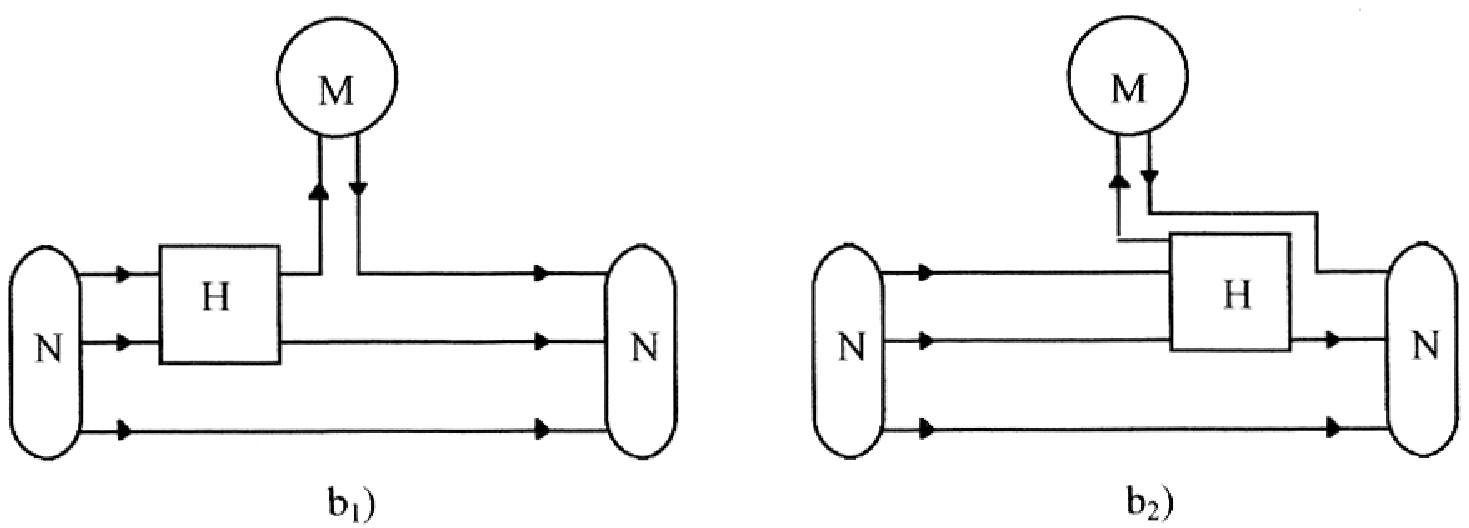,width=115mm}}
\vskip 0.5cm
\mbox{\psfig{file=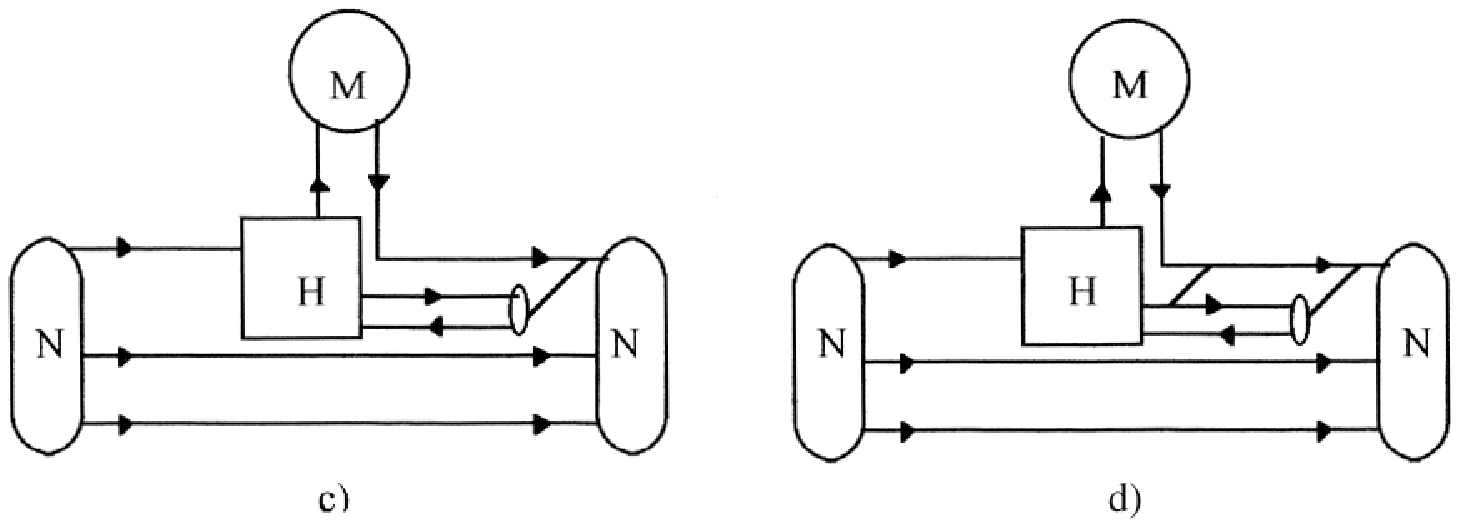,width=115mm}}
\caption{ Quark processes contributing to the pnc meson-nucleon couplings. 
Diagrams labelled a, b, c, d, arise respectively from the factorization 
approximation, parity admixture into the nucleon, first order and 
part of the second order sea quarks. Diagram c) has been corrected 
with respect to the one presented in Ref. \cite{DESP}. 
In all diagrams, the box represents the pnc quark interaction. 
Its effective character can be seen from the examination of the last 
two diagrams. By extending the box, it is possible to have it to 
incorporate that part involving gluon exchange so that to get a 
diagram a) type. On the other hand, the $\bar{q} q $ pair, which 
the gluon couples to, contains in particular a $\bar{s} s $ pair, 
colored in c), partly uncolored in d). This last diagram involves   
a sub-diagram similar to that one responsible for reducing the contribution of 
quarks to the nucleon spin. \label{c-figure}  } 
\end{center}
\end{figure}

Details on the theoretical estimates of the pnc coupling constant, $h^1_{\pi}$, 
are given here. As well known, this coupling depends on the isovector part of 
the hadronic weak interactions, which reads in the ``free'' case:
\begin{eqnarray}
H^{pnc}(\Delta T=1)= \frac{G_{F}}{\sqrt{2}} \left( \frac{1}{2} sin^2 \theta_c \,
[\bar{s} \gamma^{\mu} u \, \bar{u} \gamma_{\mu} \gamma_5 s +
\bar{s} \gamma^{\mu}\gamma_5 u \, \bar{u} \gamma_{\mu}  s -u \rightarrow d]
 \right. \nonumber \\
\left. -\frac{1}{3} sin^2\theta_{{\rm W}}  
   \bar{s} \gamma^{\mu} s \, (\bar{u} \gamma_{\mu} \gamma_5 u 
-\bar{d} \gamma_{\mu} \gamma_5 d) \right. \nonumber \\
\left.  - ( \frac{1-2sin^2\theta_{{\rm W}}}{2} )
[      \bar{s} \gamma^{\mu} s   \, 
   (\bar{u} \gamma_{\mu} \gamma_5 u -\bar{d} \gamma_{\mu} \gamma_5 d)   
    +  \bar{s} \gamma^{\mu}\gamma_5 s \,
   (\bar{u} \gamma_{\mu} u -\bar{d} \gamma_{\mu}  d)  ] \right. \nonumber \\
\left.  - \frac{1}{3} sin^2\theta_{{\rm W}}  
   ( \bar{u} \gamma^{\mu} u + \bar{d} \gamma^{\mu} d ) \, 
   (\bar{u} \gamma_{\mu} \gamma_5 u -\bar{d} \gamma_{\mu} \gamma_5 d) \right).
   \label{eq:hpnc}
\end{eqnarray}

In practice, it can be calculated from an effective interaction  which 
incorporates effects due to strong interactions (QCD). In the 
four-flavor case, this interaction involves the factor, 
\begin{equation}
 K= 1+\frac{25\,g^2(\mu)}{48 \pi^2} \;{\rm ln}(\frac{M^2_W}{\mu^2}). 
\end{equation}
The quantity, $g^2(\mu)$, is the QCD running coupling constant at the energy 
scale $\mu$. This energy scale is in principle arbitrary, but an optimal 
value often quoted in the literature is $\mu=1$\,GeV. The determination of
this coupling has varied (and improved) with time. It has dropped from the value 
1.0 used in DDH, corresponding to $K=6$, to a value 0.4, corresponding to $K=3$. 
For this value, the detail of the various contributions to the coupling constant  
$h^1_{\pi}$ estimated in DDH reads:

\begin{eqnarray} 
h^1_{\pi}= f^c_{\pi} \,(\;\; 1 \, + \,3.8 \, - \;\;\;\;\; 3.3 \;\;\;\;\;\; 
+ \;\;\;\;\;\;\; 0.5 \;\;\;\;\;\;\; + \;\;\; 3.3 \;\;\;)=2.0\, 10^{-7}. 
\nonumber \\
\bar{s} s \; + \; \bar{s} s \, + (\bar{u} u + \bar{d} d) + par. \; admixt. + \; 
fact. 
\;\;\;\;\;\;\;\;\;\;\;\;\;\;\;\;\;\;\;\;\; \label{eq:detail}
\end{eqnarray}

The sign in this expression refers to a positive sign of the strong coupling, 
$g_{\pi NN}$, appearing in Eq. 4. Otherwise, all the contributions are 
normalized to the contribution of 
the charged current, $ f^c_{\pi}= 0.38\,10^{-7}$, which was calculated first in 
\cite{MACK0} and involves strange quarks (this is remembered below the 
corresponding contribution). The other contributions in Eq. (\ref{eq:detail}) 
arise from the neutral current. The second and third contributions, like the 
first one involves sea quarks, respectively strange and non-strange. They can 
all be traced back to the diagram c of the figure. The fourth contribution 
involves ``parity-non-conservation'' in the nucleon wave function (see diagrams 
b of the figure). The last contribution corresponds to the factorization 
approximation (diagram a of the figure).

Concerning the order of magnitude, the first term in Eq. (\ref{eq:detail}) 
contains a factor ${\rm sin}^2\,\theta_c\simeq0.05$, explaining for some part 
its small contribution. It arises from the first line in Eq. (20). The second 
one essentially involves a factor $1-2 \,{\rm sin}^2\,\theta_W \simeq 0.5$, 
which makes its contribution larger than the first one. It arises from the 
second and third lines in Eq. (20). Being of the second order in gluon exchange, 
instead of the first order, the enhancement is not as large as one could naively 
expect from the comparison of the factors mentioned above. The three other terms 
in Eq. (\ref{eq:detail}) contain a factor  
$\frac{{\rm sin}^2\,\theta_W}{3} \simeq 0.08$. They only involve u and d quarks 
and arise from the fourth line in Eq. (20). While examination of factors give 
some insight on the magnitude of each term, making a precise prediction is 
difficult due to the presence of a destructive contribution. Depending on the 
prejudice on the size of the various contributions,  one can reasonably expect 
the coupling constant, $h^1_{\pi}$ to be in the following range, $(-1,\, 3) 
\times 10^{-7}$. 

It is instructive to make a detailed comparison of the above results with other 
works. Prior to DDH, Weinberg considered the contribution from neutral currents 
involving strange quarks \cite{WEIN}. The value he got was much larger than 
obtained above but was calculated in the limit $K \rightarrow \infty$. The 
finite value of K reduces his estimate from about 12 to 3.8 in Eq. 
(\ref{eq:detail}). A contribution involving sea- but non-strange quarks was also 
considered by Gari and Reid \cite{REID}. After correction for a factor -4, their 
contribution is decreased to provide a value that is accounted for by the 
negative contribution -3.3 in Eq. (\ref{eq:detail}). Dubovik and Zenkin 
\cite{DUBO} considered contributions that 
can be compared (and do compare) to the fourth and fifth numbers in this 
equation. Using QCD sum rules, Khatsimovskii \cite{KHAT} made an estimate 
which turns out to be larger than in Eq. (\ref{eq:detail}) and close 
to the ``best guess'' DDH value. In principle, it should involve some of the 
three last contributions in this equation (see comment below). Kaiser and 
Meissner \cite{KAIS}, using a quite different approach based on a chiral 
Lagrangian, calculated a contribution that involves non-strange quarks and got a 
small number for the coupling $h^1_{\pi}$ .  This one compares reasonably well 
with the sum of the three last contributions in Eq. (\ref{eq:detail}). With this 
respect the possibility that one of the contributions has a destructive 
character is important and should deserve some attention. The role of strange 
quarks, neglected in the last four works just mentioned above but taken into 
account in earlier works was reminded by Kaplan and Savage \cite{SAVA2}. Their 
estimate agrees with that one made by Weinberg and, in view of its very rough 
character, can be considered as consistent with the contribution proportional to 
3.8 in Eq. (\ref{eq:detail}). Another QCD sum rule estimate by Henley, Hwang and 
Kisslinger \cite{HENL}, always neglecting strange quarks, first turned out to be 
in agreement with what was expected from the three last numbers in Eq. 
(\ref{eq:detail}). After correction for the relative sign of two contributions 
(see errata in \cite{HENL} and \cite{HWAN}), this estimate came closer to that 
one made previously by Khatsimovskii. More recently, Meissner and Weigel 
\cite{MEIS} considered the contribution of strange quarks in a chiral model. 
Their estimate is close to that one given by the number 3.8 in Eq. 
(\ref{eq:detail}).

While most contributions considered in the literature have a counterpart in the 
DDH framework, there is a serious problem with QCD sum rule based 
calculations. One can imagine that they take into account a new contribution, 
but one would like to know which one in terms of quarks, knowing that DDH 
accounts for the simplest ones depicted in Fig. 1 (a,b,c). The other possibility 
is that something is missed by these approaches. With this respect, we would 
like to mention that a recent examination of the contribution involving sea 
quarks (Fig. 1a) led us to a contribution with a sign opposite to what is 
suggested by the analysis of hyperon decay amplitudes, used in DDH. In such a 
case, the two contributions due to non-strange quarks involving the $\bar{q} q$ 
vacuum expectation value, $<g.s.|\;\bar{q} \,q\;|g.s.>$, (3rd and 
5th terms in Eq. (\ref{eq:detail})), would add together. However, it seems that 
such an estimate misses a contribution that is required to ensure the 
conservation of an axial current coupled to the gluon field (there is some 
similarity with the calculation of an anapole moment). Enforcing this 
conservation might lead to a contribution with an opposite sign, in agreement 
with Eq. (\ref{eq:detail}). 

As mentioned above, the energy scale, $\mu$, which the K factor depends on, is 
arbitrary. Two contributions in Eq. (\ref{eq:detail}) evidence a strong 
dependence on this factor K and, through it, on the energy scale (see 
\cite{DESP}). This one, 
which determines the frontier between strong interaction effects incorporated 
separately in the effective weak interaction and in the description of hadrons, 
should not intervene in the determination of physical quantities. In the case of 
the fifth contribution to Eq. (\ref{eq:detail}) (factorization), this result is 
qualitatively achieved by cutting off the high momentum part of the effective 
interaction between quarks (as is expected), which, most probably, is equivalent 
to the effect of introducing a dependence of quark masses on the energy scale 
\cite{LEUT}. In the case of the strange quark contribution (second term in Eq. 
(\ref{eq:detail})), which vanishes in absence of strong interaction ($K=1$), 
the energy scale independence requires to introduce a strange quark-antiquark 
pair, $\bar{s} s$, in the description of hadrons. A contribution that may play 
some role in this respect is represented in Fig. 1d. It has some relationship to 
the contribution that tends to reduce that part of the nucleon spin carried by 
quarks. 

 Concerning notations, let's mention that the coupling $h^1_{\pi}$ used here is 
denoted $h^1_{\pi NN}$, or $f_{\pi}$ in other works. The notation $f_{\pi}$ is 
confusing as it represents another quantity frequently used in the field of 
strong interactions. The coupling, $h^1_{\pi}$, is also related to the quantity  
introduced in \cite{ADEL}, $H^1_{\pi}= \frac{g_{\pi NN} h^1_{\pi}}{ \sqrt{32} }= 
2.38 \, h^1_{\pi}$.


\end{document}